# Giant Bulk Electro-photovoltaic Effect in Hetero-nodal-line Systems

Xiao Jiang[1], Lei Kang[2], Jianfeng Wang[3], Bing Huang*[1,4]

[1] *Beijing Computational Science Research Center, Beijing 100193, China*

[2] *Technical Institute of Physics and Chemistry, Chinese Academy of Sciences,*

*Beijing 100190, China*

[3] *School of Physics, Beihang University, Beijing 100191, China*

[4] *Beijing Normal University, Beijing 100875, China*
E-Mail: Bing.Huang@csrc.ac.cn

**Realization of giant and continuously tunable second-order photocurrent is desired for many nonlinear optical (NLO) and optoelectronic applications, which remains to be a great challenge. Here, based on a simple two-band model, we propose a concept of bulk electro-photovoltaic effect, that is, an out-of-plane external electric-field ($E_{ext}$) can continuously tune in-plane shift current along with its sign flip in a hetero-nodal-line (HNL) system. While strong linear optical transition around the nodal-loop may potentially generate giant shift current, an $E_{ext}$ can effectively control the radius of nodal-loop, which can continuously modulate the shift-vector components inside and outside nodal-loop holding opposite signs. This concept has been demonstrated in the HNL HSnN/MoS$_2$ system using first-principles calculations. The HSnN/MoS$_2$ hetero-bilayer not only can produce giant shift current with 1~2 magnitude order larger than other reported systems, but also can realize a giant bulk electro-photovoltaic effect. Our finding opens new routes to create and manipulate giant NLO responses in 2D materials.**

*Introduction.* Nonlinear light-matter interaction plays a key role in manipulating light and matter at the nanoscale. In particular, the shift current, also known as bulk photovoltaic effect (BPVE), is a second-order nonlinear optical (NLO) process that can directly convert light into electricity, which does not require a traditional *p-n* junction and occurs only in crystals with broken inversion symmetry [1-3]. The BPVE holds great advantages in many modern physical applications, e.g., new generation of solar cells [3-6], ultra-sensitive optical detectors [7], and optically readable memristors and circuits [8]. Among all these applications, the key step relies on the realization of giant and highly tunable NLO responses, which, however, is still a great challenge [5,8-15].

In the past years, great efforts have been made to understand the mechanism of BPVE and to find the ingredients for large shift-current response [2,3,16-23]. Until now, although it is still an open question, some preliminary conclusions are made, e.g., while the strength of shift current may not depend on the intrinsic polarization of the system [3,16], it depends on the linear optical transition and shift-vector matrix elements [9-12,24,25]. Unfortunately, how to effectively tune the shift-vector matrix elements in a controllable way, which is key for tunable shift-current conductivity, is still largely unknown. Recently, several methods are proposed to realize tunable shift currents, e.g., via strain engineering (bulk piezo-photovoltaic effect [26,27] and flexo-photovoltaic effect [28]) and via ferroicity engineering [10], however, a continuous tunability in a very large photocurrent range is still unachievable.

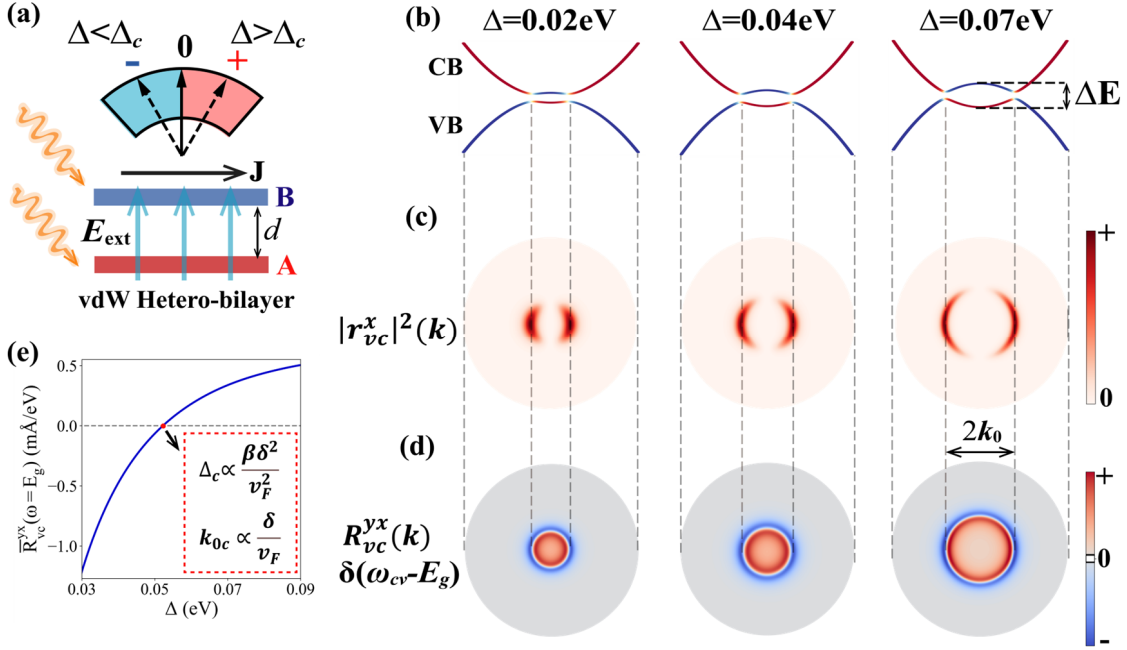

**Fig 1. Bulk electro-photovoltaic effect in an HNL system.** (a) A vdW hetero-bilayer, made of layer-A and layer-B, under an out-of-plane $E_{ext}$. (b) Evolution of the band structures with the change of onsite-energy difference between two layers (Δ) in an HNL system. Δ can be effectively tuned by an $E_{ext}$. Red (blue) band is contributed by layer A (B). ΔE measures the strength of band crossing. $k_0$ is the radius of nodal-loop. (c) and (d) are calculated $k$-resolved $|r_{vc}^x|^2$ and $R_{vc}^{yx}\delta(\omega_{cv} - E_g)$, two major components of $\sigma_{xx}^y$ around nodal-loop, with different Δ, respectively. (e) Calculated $\bar{R}_{vc}^{yx}$ around nodal-loop as a function of Δ. Parameters of $\beta = 1$ eV/Å$^2$, $\delta = 0.01$ eV, $v_F = 0.02$ eV·Å are adopted for the two-band model calculations. A small imaginary smearing factor of 0.02 eV is adopted for the numerical calculation of $\delta(\omega_{cv} - E_g)$ function. Conclusion shown here is insensitive to the different choices of parameters.

In this Letter, based on a simple two-band model, we propose that an out-of-plane external electric-field ($E_{ext}$) can continuously tune in-plane shift current along with its sign flip in a hetero-nodal-line (HNL) system. We name this effect as bulk electro-photovoltaic effect (BEPVE). The strong linear optical transition around the nodal-loop may play a key role in generating giant shift current in an HNL system. Importantly, an $E_{ext}$ can effectively control the radius of nodal-loop, which can continuously tune the shift-vector components inside and outside nodal-loop with opposite signs. Using first-principles calculations (see **Section I** [29] for details), we discover that the vdW hetero-bilayer HSnN/MoS$_2$ is an HNL system that can exhibit giant shift current (~3000 μA/V$^2$) around the nodal-loop region. Surprisingly, a relatively small $E_{ext}$ enables a giant BEPVE with a continuous current change from ~-6000 to ~6000 μA/V$^2$, which is very difficult to be realized in previously reported BPVE systems.

***Concept of BEPVE.*** Under electric field $E^b(\omega)$ at frequency $\omega$ and linearly polarized in the $b$ direction, the shift current density $J$ is a second-order response, i.e., $J^a = \sigma_{bb}^a(\omega)E^b(\omega)E^b(-\omega)$, where $a, b$ are Cartesian indices and the shift-current conductivity $\sigma_{bb}^a$ [2] is given by

$$\sigma_{bb}^a(0;\omega,-\omega) = \frac{\pi e^3}{\hbar^2}\int \frac{d^3k}{8\pi^3}\sum_{nm} f_{nm}|r_{nm}^b|^2 R_{nm}^{ab}\delta(\omega_{mn}-\omega)\ (1).$$

In **Eq.(1)**, the $r_{nm}^b$ is optical-transition matrix element, which is related to the inter-band velocity matrix element $v_{nm}$ via $r_{nm} = \frac{v_{nm}}{i\omega_{nm}}$ ($m \neq n$). $R_{nm}^{ab}$ is the shift vector that can be interpreted as the position change of a wave packet during its transition from $n$th band to $m$th band [2,16]. We can integrate $R_{mn}^{ab}$ over entire Brillouin zone (BZ) to obtain the aggregate shift vector $\bar{R}^{ab}$ [16], i.e., $\bar{R}^{ab}(\omega) = \Omega \int \frac{d^3k}{8\pi^3}\sum_{nm} f_{nm} R_{nm}^{ab}\delta(\omega_{mn}-\omega)$, where $\Omega$ is the volume of the unit cell. $\bar{R}^{ab}$ provides qualitative information about the strength and sign of shift vector [16,24,25].

We consider a general 2×2 Hamiltonian to simulate the band structure of a HNL system:
$$H(k) = \sum_i \sigma_i f_i = \delta\sigma_x + v_F k_y \sigma_y + (\beta k^2 - \Delta)\sigma_z,\quad (2)$$
where $\sigma_i = \sigma_x, \sigma_y, \sigma_z$ are the Pauli matrices, $\delta$ is the interlayer coupling strength, and $\Delta$ is on-site energy difference between two layers. Without losing generality, we take the linear term $v_F k_y$ to be in the $y$ direction, which breaks the inversion symmetry and gives rise to the nonlinear shift current. The energies of conduction band (CB) and valence band (VB) are given by $E_c = \epsilon$ and $E_v = -\epsilon$, respectively, and $\epsilon = (\sum_i f_i f_i)^{1/2}$. Subscripts $v$ and $c$ indicate the VB and CB, respectively. The radius of nodal-loop in momentum space is $k_0 = \sqrt{\Delta/\beta}$. On the nodal-loop, the energy difference between $E_c$ and $E_v$ is
$$E_c - E_v = 2\sqrt{\delta^2 + v_F^2 k_y^2},\quad (3)$$
and the bandgap of the HNL system is $E_g = 2\delta$. As indicated in **Eq. (3)**, $v_F$ determines the anisotropy for the energy difference between CB and VB on the nodal-loop.

There are two independent non-vanishing components of $R_{vc}^{ab}$ (i.e., $R_{vc}^{yx}$ and $R_{vc}^{yy}$) in the HNL system described by **Eq. (2)**. Based on the formula derivation (see **Section II** [29]), $R_{vc}^{yx}$ can be given as

$$R_{vc}^{yx} = \frac{1}{\frac{\delta}{v_F} + \frac{v_F}{\delta}k_y^2} \frac{\Delta - \beta k^2}{\sqrt{(\beta k^2 - \Delta)^2 + \delta^2 + v_F^2 k_y^2}}. \quad (4)$$

The expression of $R_{vc}^{yy}$ is complex. When $k_y = 0$, it can be simplified as $R_{vc}^{yy} = \frac{2\beta\delta}{v_F\sqrt{\delta^2 + (\beta k^2 - \Delta)^2}}$.

For the vdW HNL system [**Fig. 1(a)**], its band structures under different $\Delta$ is shown in **Fig. 1(b)**. Obviously, the larger the $\Delta$, the larger the band crossing $\Delta E$ ($\Delta E = 2\sqrt{\Delta^2 + \delta^2}$). It is interesting to explore the $|r_{vc}^x|^2$ and $R_{vc}^{yx}\delta(\omega_{cv} - E_g)$, two key ingredients of $\sigma_{xx}^y$ around nodal-loop, as a function of $\Delta$. As shown in **Fig. 1(c)**, the calculated $|r_{vc}^x|^2$ is mainly concentrated around the nodal-loop region, consistent with the fact that $v_{nm}$ ($\omega_{nm}$) has the maximum (minimum) value around the nodal-loop (**Fig. S2** [29]). Meanwhile, compared to a nodal-point (NP) system, the 2D HNL can generate much stronger joint density of states (JDOS) [$\int \delta(\omega_{cv} - E_g)$] around nodal-loop, which increases as the $k_0$ increases [30,31]. Therefore, the ultra-large linear optical transition [$\int |r_{vc}^x|^2 \delta(\omega_{cv} - E_g)$] may potentially induce giant $\sigma_{xx}^y$ around nodal-loop.

In an HNL system, $\bar{R}_{vc}$ can be well separated into two parts, i.e., $\bar{R}_{vc} = \bar{R}_{vc}^{A \to B} + \bar{R}_{vc}^{B \to A}$, in which $\bar{R}_{vc}^{A \to B}$ and $\bar{R}_{vc}^{B \to A}$ are contributed by the optical transition inside nodal-loop ($k < k_0$) and outside nodal-loop ($k > k_0$), respectively. As indicated in **Eq. (4)**, $R_{vc}^{yx}(k) = 0$ on the nodal-loop ($k = k_0$); the signs of $R_{vc}^{A \to B}$ and $R_{vc}^{B \to A}$ for the $R_{vc}^{yx}$ component are always opposite. As shown in **Fig. 1(d)**, by tuning $\Delta$ ($k_0$), it is possible to smoothly tune the values of $R_{vc}^{A \to B}\delta(\omega_{cv} - E_g)$ and $R_{vc}^{B \to A}\delta(\omega_{cv} - E_g)$, realizing the smooth change of $\bar{R}_{vc}^{yx}$ around the nodal-loop region. However, this sign flip across the nodal-loop will not happen for $R_{vc}^{yy}$ (**Fig. S3** [29]). **Fig. 1(e)** shows the calculated $\bar{R}_{vc}^{yx}$ as a function of $\Delta$. Indeed, with the increase of $\Delta$, $\bar{R}_{vc}^{yx}$ will gradually change and eventually flip its sign at a critical $\Delta$ ($\Delta_c$) or critical $k_0$ ($k_{0c}$). Based on formula derivation with some approximations (see **Section II** [29]), we can estimate $\Delta_c$ and $k_{0c}$ as

$$\Delta_c \propto \frac{\beta\delta^2}{v_F^2}, \qquad k_{0c} \propto \frac{\delta}{v_F}. \quad (5)$$

Interestingly, the $k_{0c}$ and $\Delta_c$ are strongly related to the $\delta$, $\beta$, and $v_F$ values, which are variable in different HNL materials. In practice, the $\Delta$ can be effectively tuned by the external $E_{\text{ext}}$, i.e., when an $E_{\text{ext}}$ is applied, the $\Delta$ is changed by $\Delta V = E_{\text{ext}} \cdot d$ [$d$ is the interlayer distance, **Fig. 1(a)**]. Based on this analysis, we can propose a novel concept of electric-field-tunable shift-vector (shift-current conductivity) in an HNL system, as illustrated in **Fig. 1(a)**, named as BEPVE.

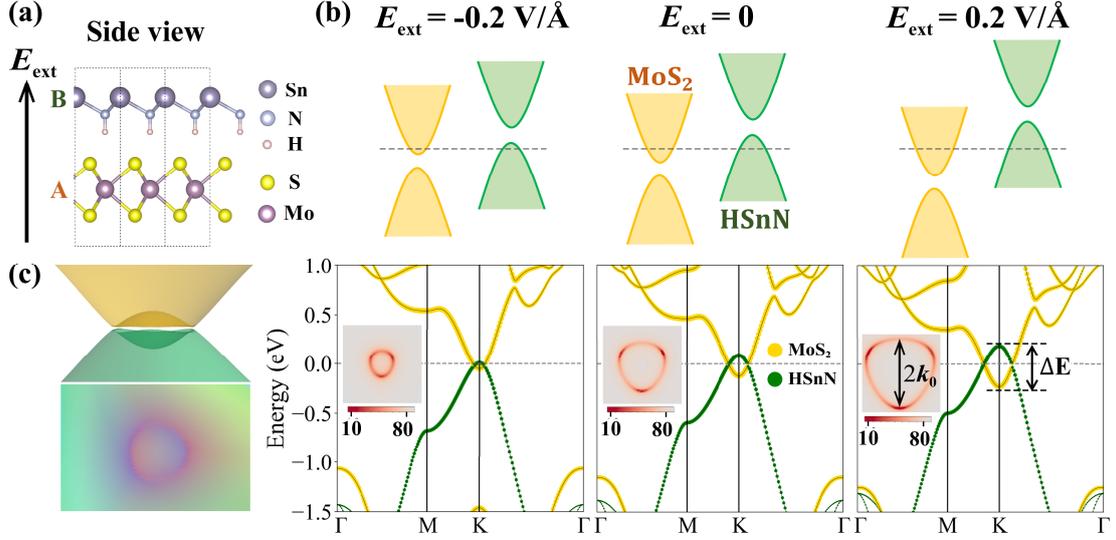

**Fig 2. HNL in HSnN/MoS$_2$.** (a) Side view of vdW hetero-bilayer HSnN/MoS$_2$. Unit cells are marked as dashed-lines. (b) Schematic illustration (up-panel) and projected band structures (down-panel) of HSnN/MoS$_2$ under $E_{ext}$, in which the direction of $E_{ext}$ is marked in (a). Insets of (b): Nonuniform anti-crossing bandgaps on the nodal-loop (unit: meV). (c) Side (up-panel) and top (down-panel) views of 3D band structure of HSnN/MoS$_2$ around HNL ($E_{ext}$=0 V/Å). Fermi level is set to zero.

***HNL in HSnN/MoS$_2$.*** Both monolayer $MX_2$ ($M$=Mo, W; $X$=S, Se) and $A$Sn$X$ ($A$=Na, H; $X$=N, P) were proposed to have strong NLO responses [10,12,32-36]. Here we focus on the vdW hetero-bilayer formed by MoS$_2$ and HSnN. The HSnN is both dynamically and thermodynamically stable (**Fig. S4** [29]). The in-plane lattice mismatch between MoS$_2$ and HSnN is relatively small (~2%). The ground-state configuration is AB-stacking with a P3m1 space group (**Fig. 2(a)** and **Fig. S5** [29]). **Figure 2(b)** shows the band structures of HSnN/MoS$_2$. The bottom of CB in MoS$_2$ is contributed by Mo $3d_{z^2}$ orbital, while the top of VB in HSnN is contributed by Sn $5p_z$ orbital (**Fig. S6** [29]). When $E_{ext}$=0 V/Å, these two band-edge states can cross with each other, resulting in ΔE=0.21 eV. As shown in **Fig. 2(c)**, the 3D band structure indicates that an HNL is formed in the entire BZ around Fermi level. The nodal-loop exhibits a $C_{3v}$-like symmetry. Interestingly, the orbital hybridization between Mo $3d_{z^2}$ and Sn $5p_z$ can induce a small anti-crossing bandgap, which is nonuniform and in the range of 10~80 meV on the nodal-loop [inset, **Fig. 2(b)**]. The $v_F$ of the material plays an important role in determining the nonuniformity of anti-crossing bandgap [**Eq. (3)**] and the $\Delta_c/k_{0c}$ [**Eq. (5)**]. The band structure around Fermi level has small changes under hybrid-functional calculations or with the inclusion of spin-orbital coupling (SOC) effect (**Fig. S7** [29]).

As shown in **Fig. 1(b)**, an out-of-plane $E_{ext}$ can adjust the ΔE between MoS$_2$ and HSnN. Indeed, as shown in **Fig. 2(b)**, a positive (negative) $E_{ext}$ can upshift (downshift) the energy level of HSnN with respect to MoS$_2$, increasing (decreasing) ΔE. Although the shape of nodal-loop is not circular, we can approximately estimate the $k_0$ in **Fig. 2(b)**. With $E_{ext}$ gradually increasing from -0.2 to 0.2 V/Å, $k_0$ can increase from 0.14 Å$^{-1}$ to 0.41 Å$^{-1}$. Overall, the HSnN/MoS$_2$ is a good HNL system with tunable nodal-loop, which is suitable to test our proposed BEPVE.

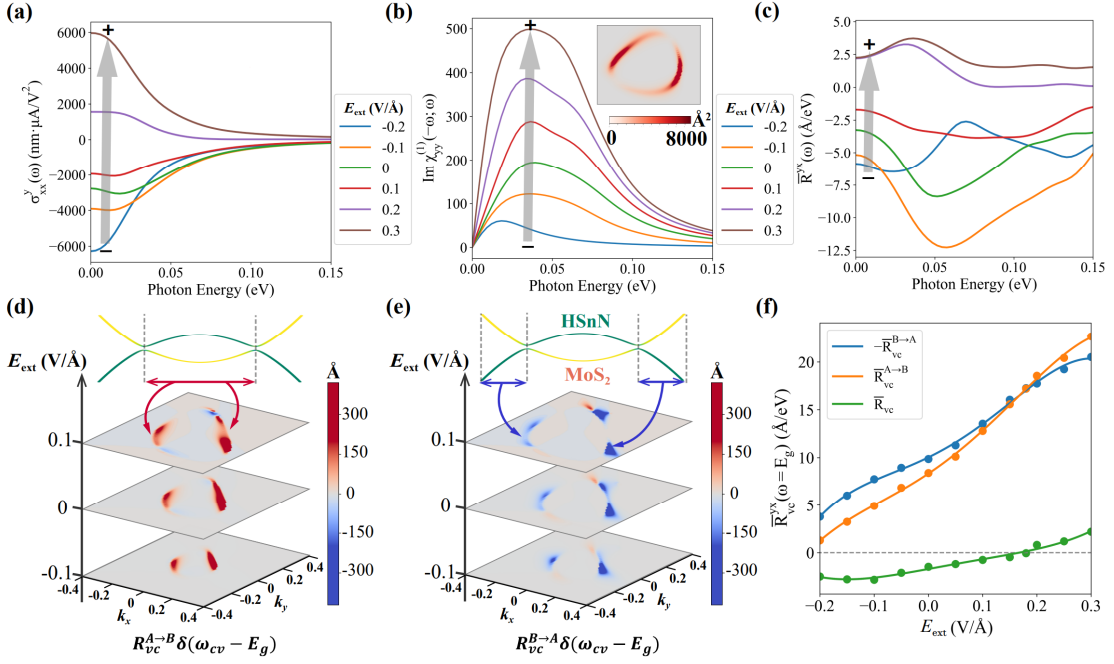

**Fig 3. $E_{ext}$-tunable $\sigma_{xx}^y$ around nodal-loop.** (a) Shift-current conductivities $\sigma_{xx}^y$, (b) linear absorption spectra $\text{Im}\chi_{xx}^{(1)}$, and (c) aggregate shift vectors $\bar{R}^{yx}$ in HSnN/MoS$_2$ as a function of photon energy under different $E_{ext}$. Inset of (b): $k$-resolved $|r_{vc}^x|^2$ between two HNL bands. (d) and (e) are $k$-resolved $R_{vc}^{A\to B}\delta(\omega_{cv}-E_g)$ and $R_{vc}^{B\to A}\delta(\omega_{cv}-E_g)$ between these two HNL bands around nodal-loop, respectively. (f) $\bar{R}_{vc}^{A\to B}$, $\bar{R}_{vc}^{B\to A}$ and $\bar{R}_{vc}^{yx}$ between two HNL bands around nodal-loop as a function of $E_{ext}$. Here, a small imaginary smearing factor of 0.02 eV is adopted for the numerical calculation of $\delta(\omega_{cv}-E_g)$ function. A (B) → B (A) means MoS$_2$ (HSnN) → HSnN (MoS$_2$).

*Giant BPVE around Nodal-loop.* Based on the symmetry restriction, there is only one (four) independent non-vanishing σ component in monolayer MoS$_2$ (HSnN). We focus on the in-plane $\sigma_{xx}^y$ ($\sigma_{xx}^y = \sigma_{xy}^x = -\sigma_{yy}^y$) component, which is much more useful than that of out-of-plane ones for device applications and also consistent with our model in **Fig. 1**. The calculated $\sigma_{xx}^y$ spectra of monolayer MoS$_2$ and HSnN are shown in **Fig. S8** [29]. Overall, the peak value at > 2 eV of HSnN (~80 nm·µA/V$^2$) is twice larger than that of MoS$_2$ (~40 nm·µA/V$^2$). Interestingly, as shown in **Fig. 3(a)**, when the hetero-bilayer is formed, besides the same peak spectra at high photon-energy as individual monolayers, a new huge peak emerges around the nodal-loop, which decreases gradually as the photon-energy increases. This new peak value is as huge as ~-3000 nm·µA/V$^2$, remarkably higher than existing materials.

Next, we consider the $E_{ext}$ effect. It is not surprising that the $\sigma_{xx}^y$ of both MoS$_2$ and HSnN are insensitive to the $E_{ext}$ (**Fig. S9** [29]), because their electronic structures are insensitive to the $E_{ext}$. Remarkably, after forming HNL, the $\sigma_{xx}^y$ of HSnN/MoS$_2$ strongly varies with the $E_{ext}$. As shown in **Fig. 3(a)** and **Fig. S10(a)** [29], when $E_{ext}$ gradually changes from -0.2 to ~0.18 V/Å, the peak around nodal-loop region gradually decreases from ~-6280 to ~0 nm·µA/V$^2$. When $E_{ext}$ further increases from ~0.18 to 0.3 V/Å, the peak gradually increases from ~0 to ~6000

nm·µA/V$^2$, along with the sign switching. This conclusion is not changed when the SOC effect is included (**Fig. S10(b)** [29]). Clearly, the formation of HNL plays a key role in forming (i) huge $\sigma_{xx}^y$ peak and (ii) $E_{\text{ext}}$-tunable $\sigma_{xx}^y$ peak around nodal-loop region in HSnN/MoS$_2$.

It is interesting to further explore the origin of (i)-(ii) in HSnN/MoS$_2$. As shown in **Fig. 3(b)**, the ultra-strong linear optical absorption spectrum (Im$\chi_{xx}^{(1)}$), originated from the large $|r_{vc}^x|^2$ [inset of **Fig. 3(b)**] and HNL-induced large JDOS [$\int \delta(\omega_{cv} - E_g)$], could be the key to generating huge $\sigma_{xx}^y$. when $E_{\text{ext}}$=0 V/Å, the Im$\chi_{xx}^{(1)}$ is peaked around nodal-loop with a large value of ~195, which is ~22 (~18) times larger than that of monolayer HSnN (MoS$_2$) (**Fig. S4** [29]). The $k$-resolved $|r_{vc}^x|^2$ is always concentrated around the nodal-loop under different $E_{\text{ext}}$ (**Fig. S11**), agreeing with **Fig. 1(c)**. When the $E_{\text{ext}}$ gradually changes from -0.2 to 0.3 V/Å, the increased band inversion can gradually increase the $k_0$, giving rise to the increased Im$\chi_{xx}^{(1)}$.

Usually, the stronger the linear absorption, the larger the NLO response [3,9,17]. Surprisingly, our case does not follow this common rule, indicating the unusual role of shift vector. As understood in **Fig. 1(e)**, the sign switch of $\sigma_{xx}^y$ may be related to the sign change of $\bar{R}^{yx}$ under $E_{\text{ext}}$. Indeed, as shown in **Fig. 3(c)**, the trend of $\bar{R}^{yx}$, including its sign, around the nodal-loop region matches the trend of $\sigma_{xx}^y$ under different $E_{\text{ext}}$, regardless that the overall curvature of $\bar{R}^{yx}$ is quite different from that of $\sigma_{xx}^y$ [16].

We have further calculated the $k$-resolved $R_{vc}^{yx} \delta(\omega_{cv} - E_g)$ between these two HNL bands. The $R_{vc}$ can be projected into two parts, i.e., $R_{vc}^{A \to B}$ and $R_{vc}^{B \to A}$. As shown in **Fig. 3(d)** [**Fig. 3(e)**], the $R_{vc}^{A \to B} \delta(\omega_{cv} - E_g)$ [$R_{vc}^{B \to A} \delta(\omega_{cv} - E_g)$] represents the shift vector contributed by the optical transition from MoS$_2$ [HSnN] to HSnN [MoS$_2$] around nodal-loop, which is concentrated inside [outside] the nodal-loop in the $k$-space, agreeing with **Fig. 1(d)**. On the nodal-loop, $R_{vc}^{yx}$ is zero. With the $E_{\text{ext}}$ changes from negative to positive, the $k_0$ of nodal-loop increases, which can further manipulate the $k$-space distribution of $R_{vc}^{A \to B} \delta(\omega_{cv} - E_g)$ and $R_{vc}^{B \to A} \delta(\omega_{cv} - E_g)$. The $\sigma_{xx}^y$ resembles a similar $k$-space distribution to $R_{vc}^{yx} \delta(\omega_{cv} - E_g)$, concentrated around nodal-loop under different $E_{\text{ext}}$ (**Fig. S11** [29]). The calculated $\bar{R}_{vc}^{A \to B}$ and $\bar{R}_{vc}^{B \to A}$ are shown in **Fig. 3(f)**, exhibiting two major features: (1) the signs of $\bar{R}_{vc}^{A \to B}$ and $\bar{R}_{vc}^{B \to A}$ are always opposite, and (2) the slope of $|\bar{R}_{vc}^{A \to B}|$ is larger than that of $|\bar{R}_{vc}^{B \to A}|$ under $E_{\text{ext}}$. Therefore, the $\bar{R}_{vc}^{yx} = \bar{R}_{vc}^{A \to B} + \bar{R}_{vc}^{B \to A}$ can be gradually changed as a function of $E_{\text{ext}}$, agreeing with our model calculation in **Fig. 1(e)**. Interestingly, the sign of $\bar{R}_{vc}^{yx}$ are enforced to flip at a critical $E_{\text{ext}}$ of ~0.18 V/Å, so as to the $\sigma_{xx}^y$ (**Fig. S10(a)** [29]). Now, we can conclude that while the strong linear optical transition can generate giant $\sigma_{xx}^y$ peak around the nodal-loop, the $E_{\text{ext}}$-dependent $\bar{R}_{vc}^{yx}$ plays a key role in manipulating this $\sigma_{xx}^y$ peak, eventually realizing giant BEPVE.

It is interesting to further understand the material-dependent sign flip of $\bar{R}_{vc}^{yx}$ ($\sigma_{xx}^y$). Therefore, we further calculate other three HSnN/$MX_2$ ($M$=Mo, W; $X$=S, Se) materials. First, following a similar calculation procedure for HSnN/MoS$_2$, we can determine the critical $E_{ext}$ for sign flip of $\sigma_{xx}^y$ in different HSnN/$MX_2$ (**Fig. S12** [29]). Second, based on **Eq. (3)**, the values of $\delta$, $v_F$ and $k_{0c}$ in these materials are further estimated in the electronic structures under their critical $E_{ext}$ (**Table S1** [29]). Finally, as shown in **Fig. 4(a)**, the $k_{0c}$ is almost linear dependent on $\delta/v_F$ in these four systems, agreeing well with **Eq. (5)**. We note the small deviation may be due to the approximation on estimating $k_{0c}$ in HSnN/$MX_2$, i.e., the shape of nodal-loop in HSnN/$MX_2$ is not ideally circular [**Fig. 2(b)**], differing from the one in two-band model [**Fig. 1(d)**]. In practice, via controlling the $\delta$, $\beta$, and $v_F$ values in different HNL systems, we may realize the predesigned shift-current switch under different $E_{ext}$, which may be used for different device functions.

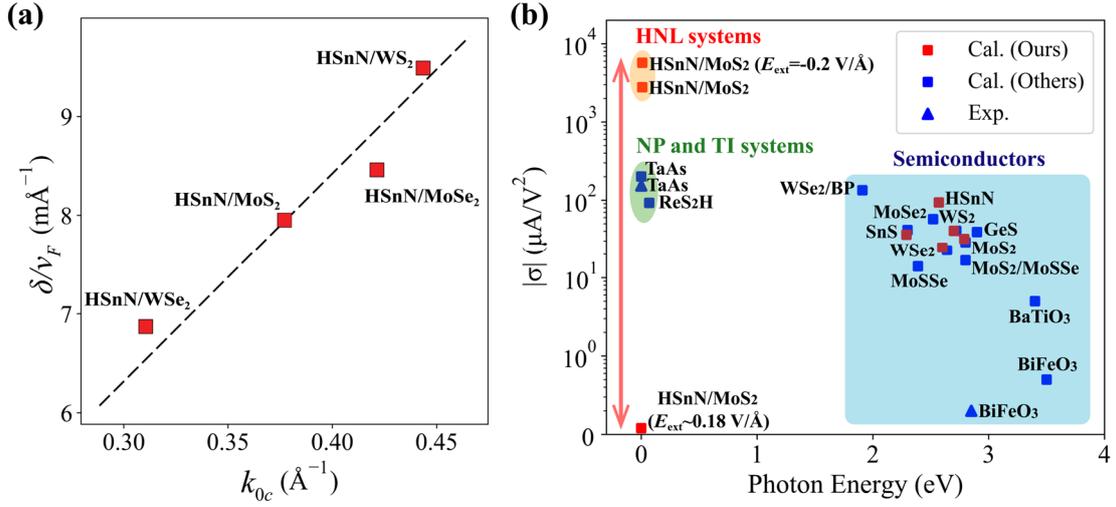

**Fig 4.** (a) Relation between calculated $\delta/v_F$ and $k_{0c}$ in four HSnN/$MX_2$ ($M$=Mo, W; $X$=S, Se) systems, which shows a nearly linear dependence. (b) Comparison of |σ| between heterobilayer HSnN/MoS$_2$ and other widely studied BPVE materials in the literature [7,9-11,13,16,37-39].

***Discussion.*** It is interesting to compare the σ of HSnN/MoS$_2$ with other widely studied BPVE materials. As shown in **Fig. 4(b)**, we have collected the peak values in the largest σ component around bandgap in a couple of 2D and 3D BPVE systems reported in the literature [7,9-11,13,16,37-39]. These values have been carefully converted into the same unit of μA/V$^2$ for direct comparison. Among these materials, we have recalculated four of them (**Fig. S13** [29]), i.e., MoS$_2$, SnS, WS$_2$, and WSe$_2$, using similar computational parameters as the literature [10], which produce similar results, validating our computational method. Interestingly, these BPVE materials in **Fig. 4(b)** can be roughly divided into three categories: (i) common semiconductors with sizable bandgap, e.g., $|\sigma_{xx}^z|$ ~5 μA/V$^2$ for BaTiO$_3$ [16] and $|\sigma_{yy}^y|$ ~40 μA/V$^2$ for monolayer MoS$_2$ [10]; (ii) NP and topological insulator (TI) systems with tiny bandgap, e.g., $|\sigma_{xz}^x|$~200 μA/V$^2$ for TaAs [7,37,40] and $|\sigma_{yy}^y|$~90 μA/V$^2$ for ReS$_2$H [9]; (iii) our proposed HNL system with tiny bandgap, e.g., $|\sigma_{yy}^y|$~2800 μA/V$^2$ and ~6280 μA/V$^2$ for HSnN/MoS$_2$ under an $E_{ext}$ of 0 and -0.2 V/Å, respectively. Overall, the |σ| of category (iii) is one [two] order of magnitude larger than that of category (ii) [(i)], indicating that HNL system is suitable to generate giant |σ|. Most importantly, besides of the sign flip, as marked by the arrow in **Fig.**

**4(b)**, the $|\sigma_{xx}^{y}|$ value of HSnN/MoS$_2$ can be hugely and continuously tuned from 0 to ~6000 μA/V$^2$ under relatively small $E_{\text{ext}}$.

In summary, we provide a new route to create and continuously manipulate giant shift currents in 2D HNL systems via a concept of bulk electro-photovoltaic effect. More generally, our idea may be applicable to other NLO phenomena, e.g., second harmonic generation and second-order spin photocurrent, in which the smooth control of shift vector is always the key for manipulating the overall NLO responses.

**Acknowledgment:** We thank Dr. Haowei Chen at Tsinghua University for helpful discussion. This work is supported by the National Key Research and Development of China (Grant No. 2022YFA1402400), NSFC (Grants Nos. 12088101 and 12174404), NSAF (Grant No. U2230402), and Open Research Fund Program of the State Key Laboratory of Low-Dimensional Quantum Physics. Calculations were done in Tianhe-JK cluster at CSRC.